\documentclass[aps,preprintnumbers,amsmath,amssymb,superscriptaddress,groupedaddress,floatfix,nofootinbib]{revtex4}
\usepackage{graphicx}
\usepackage[normalem]{ulem}
\usepackage{amsmath} 
\usepackage{verbatim} 
\usepackage{mathrsfs}
\usepackage{color}
\usepackage{amsfonts}
\usepackage{graphics}
\usepackage{epsfig}
\usepackage{hyperref}
\usepackage{subcaption}

\newcommand{\zobs}{z_{\rm obs}}

\newcommand{\zdot}{\dot z}
\newcommand{\shat}{\hat{s}}  
\newcommand{\Hub}{H}
\newcommand{\Hzero}{H_0}
\newcommand{\dzdot}{\delta\dot z}

\begin{document}

\title{Expected redshift drift for tilted observers}

\author{Franco R. de Pedro}
\email{fdepedro@df.uba.ar}
\affiliation{Facultad de Ciencias Exactas y Naturales,
Departamento de Física, Universidad de Buenos Aires. Buenos Aires, Argentina.}

\author{Gabriel R. Bengochea}
\email{gabriel@iafe.uba.ar}
\affiliation{Instituto de Astronom\'{\i}a y F\'{\i}sica del Espacio (IAFE),
CONICET--Universidad de Buenos Aires, (1428) Buenos Aires, Argentina.}

\begin{abstract}

Redshift drift is usually discussed for observers comoving with the cosmological background, but realistic observations are made by observers with nonzero peculiar motion. In this work, we calculate the expected redshift drift for tilted observers within the covariant 1+3 formalism. Starting from the exact redshift measured in the tilted frame, we derive the corresponding drift as an FLRW background term plus a directional correction driven by the observer's peculiar kinematics, encoded through peculiar expansion, projected shear, and projected acceleration along the line of sight. We analyse first an Einstein--de Sitter (EdS) background, which isolates the purely kinematic effect of tilt in the absence of background acceleration, and then extend the calculation to $\Lambda$CDM in order to quantify how the same anisotropic corrections deform the standard drift signal in the concordance model. 

\end{abstract}

\maketitle

\section{Introduction}
\label{sec_intro}

Since the late 1990s, observational cosmology has consolidated a broad consensus around the standard cosmological model $\Lambda$CDM, based on a homogeneous and isotropic Friedmann--Lemaître--Robertson--Walker (FLRW) background plus small Gaussian perturbations, which successfully describes the cosmic microwave background (CMB), baryon acoustic oscillations (BAO), primordial nucleosynthesis (BBN), and the growth of large-scale structure \cite{Ellis2012,TCM2008,Turner2022}. In this framework, the cosmic energy budget is modeled in terms of baryons, cold dark matter (CDM), and dark energy, usually represented by a cosmological constant $\Lambda$ with equation of state $p=w\rho$ and $w=-1$. The late-time acceleration, originally inferred from Type Ia supernovae, is then interpreted as evidence for this dominant negative-pressure component, whose present fraction is constrained to be around $70\%$ by CMB, BAO, and BBN observations \cite{Planck:2018Overview,Planck:2018CosmologicalParameters,PDG:2024CosmologicalParameters,Cortes2026,PDG:2024BBN}.

At the same time, several tensions and anomalies motivate the exploration of alternative or complementary descriptions \cite{Perivolaropoulos:2022,Abdalla:2022,DiValentino:2022,DESI:2024mwx,Aluri2022}. These tensions do not, in themselves, constitute a refutation of the standard cosmological model, but they do justify the exploration of conceptual frameworks where certain kinematic and observer-selection effects are treated in a more refined manner. In particular, non-negligible peculiar motions on scales of several hundred $\mathrm{Mpc}$ suggest that the real universe, as seen by a specific observer, may depart from the strictly FLRW idealization. While inhomogeneities are usually treated as perturbations around a homogeneous background and observers are idealized as comoving with it, real observers are embedded in nonlinear structures (galaxies, clusters, voids) and participate in substantial peculiar motions with respect to the idealized Hubble flow. From a kinematical point of view, the velocity field of galaxies can be decomposed into a part proportional to the homogeneous Hubble expansion and a residual part induced by gravitational inhomogeneities. The residual part of the galaxy velocity field then gives rise to large-scale coherent motions, or \emph{bulk flows}, and the identification between the real observer and the comoving background frame has been increasingly challenged by ever more precise measurements of peculiar velocities.

One of the most discussed indications of large-scale coherence came from the kinetic Sunyaev--Zel'dovich effect in galaxy clusters: \cite{Kashlinsky2008} reported a flow of order $600$--$1000\,\mathrm{km\,s^{-1}}$ extending to $\sim 300\,h^{-1}\,\mathrm{Mpc}$ or beyond. Although the statistical robustness of this so-called \emph{dark flow} remains debated, with both critical reanalyses and independent constraints \cite{Keisler:2009,Planck:2014XIII} as well as methodological defenses and claims of a consistent signal \cite{Kashlinsky:2010BulkFlow,AtrioBarandela:2013PlanckBF}, it helped sharpen the possibility that the local volume might be embedded in a large-amplitude coherent flow.

Complementary evidence comes from peculiar-velocity catalogues such as Cosmicflows-2/3/4, 6dFGSv, and 2MTF \cite{Tully:2013,Tully:2016,Tully:2023,Campbell:2014,Springob:2014,Masters:2008,Hong:2012,Hong:2019}, which have been used to reconstruct the velocity field out to several hundred $h^{-1}\,\mathrm{Mpc}$. These analyses find coherent bulk flows on scales of order $50$--$150\,h^{-1}\,\mathrm{Mpc}$ whose amplitudes appear in some cases somewhat larger than expected in $\Lambda$CDM simulations \cite{Watkins2009,Watkins2015}. More recent Cosmicflows-4 analyses further emphasize the amplitude and statistical robustness of such large-scale flows, suggesting a possible tension with $\Lambda$CDM when CMB-derived parameters are adopted \cite{Watkins2023}. In particular, \cite{Watkins2025} decomposed the observed velocity field into an ``internal'' component sourced within $\sim 200\,h^{-1}\,\mathrm{Mpc}$ and an ``external'' one sourced beyond that volume, concluding that the coherent flow on that scale is dominated by external sources and is highly unlikely within the $\Lambda$CDM framework. This result reinforces the idea that our local cosmic environment may be embedded in a large-scale peculiar flow whose detailed understanding is crucial both for tests of isotropy and for the interpretation of cosmological parameters inferred from local observations. Recent reconstructions also suggest the presence of locally expanding and contracting regions around the Local Group; in particular, studies such as \cite{Pasten2024}, based on velocity-field reconstructions from galaxy catalogues and approximations for the divergence of the flow, are beginning to quantify the divergence of the local flow and its relevance for alternative cosmologies. This type of analysis suggests that the observer may reside in a region whose flow has negative divergence, i.e. is mildly contracting.

In this context, the relativistic framework known as the \emph{tilted universe}\footnote{In relativity, cosmological models equipped with two families of observers in relative motion are usually referred to as \emph{``tilted''} models \cite{Collins65}.} has been developed to study systematically the impact of large-scale peculiar motions on cosmological observations, e.g. \cite{Tsagas:2010MNRAS,Tsagas2011,Tsagas2012,Tsagas2015,Tsagas2021NvsR,Tsagas2022,Tsagas:2024zrx,Miliou2024,Tsagas:2025nic,Tsagas2025rev}. Its central premise is that real observers, embedded in peculiar flows, do not exactly coincide with the fundamental observers of the homogeneous and isotropic Hubble flow. As a consequence, quantities such as the deceleration parameter, the local expansion rate, or the isotropy of the sky may be interpreted differently in relatively moving frames. In particular, even on a dust FLRW background (e.g. Einstein--de Sitter), observers within mildly contracting peculiar flows may infer a negative local deceleration parameter, i.e. an \emph{apparently accelerated} expansion \cite{Tsagas2011,Tsagas2015,Tsagas2022}. In this reading, such acceleration would be an artifact of relative motion, rather than necessarily a signal of a new exotic component in the energy content of the universe.

A characteristic phenomenological feature of this scenario is that the deceleration parameter measured by a tilted observer acquires directional corrections from the divergence and shear of the peculiar flow, while the peculiar velocity itself selects a preferred axis \cite{Tsagas2011}. For small shear, this naturally leads to a Doppler-type dipole in the angular distribution of the deceleration parameter and of the local Hubble parameter, expected to decay with redshift and to be approximately aligned with the CMB dipole \cite{Tsagas2011,Tsagas2015,Tsagas2022,Tsagas:2025nic}. Observationally, dipolar anisotropies in the local expansion have been reported by several recent analyses: \cite{Migkas2021} found a $>5\sigma$ dipole in $H_0$, possibly associated with a \emph{bulk flow} of $\sim 900\,\mathrm{km\,s^{-1}}$ out to $\sim 500\,\mathrm{Mpc}$; \cite{Luongo2022} found similar sky variations using QSOs and GRBs; and \cite{Secrest2021} reported a large quasar dipole aligned with the CMB dipole and in $4.9\sigma$ tension with standard $\Lambda$CDM. These results provide a nontrivial observational motivation for frameworks in which relative-motion effects leave anisotropic imprints on cosmological observables.

Progress has also been made in confronting the tilted scenario more directly with data. An important milestone is the work of \cite{Asvesta2022}, which provided the first quantitative constraints on the deceleration parameter in a \emph{tilted universe} using Type Ia supernovae, showing that a \emph{tilted} Einstein--de Sitter model can fit the data with a goodness of fit comparable to that of $\Lambda$CDM, at least at the phenomenological level. More recently, \cite{Sah2024} analyzed anisotropies in the Pantheon+ sample, searching for dipoles in both $H_0$ and $q$ in the heliocentric, CMB, and Local Group frames. They reported a significant dipole in $H_0$ over $0.023<z<0.15$ (with a direction approximately aligned with the CMB dipole) and, more importantly, a dipole in $q$ at the $4$--$6\sigma$ level whose amplitude decreases with redshift and tends to zero for $z\gtrsim 0.5$--$0.7$, thus qualitatively reproducing the redshift-dependent Doppler-type dipolar modulation discussed in the tilted literature \cite{Tsagas2011,Tsagas:2025nic}. They further interpreted this anisotropy as evidence that at least part of the acceleration inferred from supernovae may be a relativistic effect associated with the local bulk flow. Complementary reconstructions from catalogues such as 2M++ and Cosmicflows-4 also allow one to estimate the divergence of the local flow \cite{Watkins2023,Pasten2024}, although present uncertainties still prevent definitive conclusions. Further observational studies have also reported possible directional variations in the local expansion rate, using Pantheon+ supernovae and Tully--Fisher data from Cosmicflows-4, providing additional motivation for considering anisotropic and relative motion effects in low-redshift cosmological observables \cite{McConville2023,Boubel2024}.

These phenomenological signatures make the \emph{tilted} scenario an attractive framework for exploring how large-scale peculiar motions may bias or modulate the interpretation of cosmological observations. Nevertheless, in light of the recent debate, it is important to distinguish between two different levels of the programme. On the one hand, there is the usefulness of the tilted formalism as a kinematic and phenomenological framework for describing relative-motion effects and their impact on observables. On the other hand, there is the stronger claim that the same framework can dynamically explain anomalous bulk flows or even replace dark energy as the origin of the apparent cosmic acceleration. This stronger interpretation has recently been questioned in \cite{ClarksonMaartens2026} and subsequently replied to in \cite{Tsagas2026reply,PatliakaTsagas26}, so it is presently the subject of an open debate. We will not enter into those arguments here, however, since they do not affect the presentation or the results of the present manuscript. The analysis developed in this work relies only on the exact covariant formulation of the signal as a light-cone observable and on a phenomenological parametrization of the peculiar kinematics, but not on adopting any particular relativistic growth law for bulk flows or on attributing to them, by themselves, an effective acceleration of the background.

One of the observational tests capable of probing directly the time evolution of the cosmic expansion is the measurement of the \emph{redshift drift} \cite{Sandage1962,McVittie1962,Loeb1998}. Unlike standard distance-based probes, this observable is sensitive to the actual change of the expansion rate along the observer's past light cone. More precisely, it arises from the acceleration or deceleration of the cosmic expansion through the evolution of the Hubble parameter over cosmic time, which induces a tiny temporal variation in the redshift of distant sources. Its prediction is theory-dependent, since it depends on the underlying gravitational dynamics, the spacetime geometry, and the matter content of the universe. As a result, cosmological models with different backgrounds, different symmetries, or different mechanisms for the apparent late-time acceleration can lead to substantially different drift signals, making this effect a potentially powerful discriminator among competing scenarios. Although the signal is extremely small (\(\sim 10^{-9}\) over a decade), the next generation of facilities, such as ELT/ANDES and SKA, is expected to reach the sensitivity required for its detection, possibly by around 2035; see, for instance, \cite{Balbi2007,Haehnelt2010,Liske2008,Marconi2024}. A successful measurement would therefore open a qualitatively new observational window to test the viability of different cosmological models and mechanisms proposed to account for the observed accelerated expansion. Complementary studies have addressed related effects of cosmic inhomogeneity on the redshift drift signal, from backreaction-motivated toy models and fully relativistic EdS simulations to relativistic N-body analyses of peculiar velocity and peculiar acceleration contributions in $\Lambda$CDM-like cosmologies \cite{Koksbang19,Koksbang24,Oestreicher25}. Nonlinear relativistic corrections to the redshift drift, and their possible impact on future observations and on the interpretation of nonlinear large-scale structure scales, have also been recently studied in \cite{Bechaz2026}.

Within this context, the aim of the present work is to derive the expected cosmological redshift drift in the \emph{tilted universe} scenario and to compare it with the standard \(\Lambda\)CDM prediction. More specifically, we investigate how the relative motion of a tilted observer modifies the drift signal through directional kinematical corrections associated with the local peculiar flow. In this sense, the redshift drift provides a particularly natural light-cone observable with which to assess whether large-scale peculiar motions can leave an observable imprint distinct from that expected in the concordance model.

The paper is organized as follows. In Section~\ref{sec_2}, we develop the covariant definition of cosmological redshift, derive its \(1+3\) propagation equation, recover the standard FLRW result, and obtain the redshift measured by tilted observers in terms of a directional Doppler factor. In Section~\ref{sec_3}, we derive the corresponding redshift drift expression, separating the FLRW Sandage--Loeb contribution from the peculiar-motion correction, and then estimate the expected signal using a motivated dipolar model for both EdS and \(\Lambda\)CDM backgrounds. Finally, in Section~\ref{sec_conclusions}, we summarize our main results and discuss their implications for the tilted universe interpretation of the apparent acceleration. Throughout the manuscript we adopt the spacetime metric signature \((- ,+,+,+)\), we work in natural units with \(c=1\), and set \(\hbar=1\) when convenient in the discussion of the null wave vector and the radiation energy measured by the observer, unless otherwise indicated.

\section{Covariant Cosmological Redshift}
\label{sec_2}

In this section, we develop the covariant definition of redshift in general relativity and its implementation within the \(1+3\) covariant formalism. We first present the general definition in terms of the null wave vector of the radiation field and the 4-velocities of the emitter and the observer. We then derive the \(1+3\) propagation equation for the radiation energy measured by the observer congruence, explicitly identifying the contributions due to the expansion, shear, and acceleration of that congruence. In particular, we recover the standard FLRW result and, finally, obtain the equation for the redshift measured by \emph{tilted} observers in terms of a directional Doppler factor.

\subsection{General covariant definition of redshift}
\label{redshift_cov}
Let us begin by considering a light ray propagating from an emission event \(e\) to an observation event \(o\) along a null curve \(\gamma\). Its wave vector \(k^a\), or equivalently its 4-momentum \(p^a=\hbar k^a\), is tangent to \(\gamma\):
\begin{equation}
p^a=\hbar k^a,
\qquad
k^a\equiv\frac{dx^a}{d\lambda},
\end{equation}
where \(\lambda\) is a parameter along the ray. Given the 4-velocities of the emitter and the observer, \(u_e^a\) and \(u_o^a\) (normalized as \(u^a u_a=-1\)), the frequency and energy measured locally by an observer \(u^a\) are defined covariantly (setting $\hbar=1$) by
\begin{equation}
\label{eq:nu_def}
\nu \equiv -\,u_a k^a =E.
\end{equation}
If \(u^a\) and \(k^a\) are \emph{future-directed}, then \(-u\!\cdot\!k>0\) and \(\nu>0\).

The covariant definition of redshift is then (e.g. \cite{EllisElst1998,Ellis2012}),
\begin{equation}
\label{eq:z_cov_def}
1+z \equiv \frac{\nu_e}{\nu_o}
=\frac{\left(-u_a k^a\right)_e}{\left(-u_a k^a\right)_o}
=\frac{E_e}{E_o}.
\end{equation}

The natural language for describing observers in relative motion in general relativity is the \(1+3\) covariant formalism, e.g. \cite{TCM2008,Ellis2012}. In this formalism one introduces a timelike four-velocity field \(u^a\), which defines a foliation of spacetime into ``spatial hypersurfaces'' orthogonal to \(u^a\) and temporal curves tangent to \(u^a\). Given an observer characterized by \(u^a\), any tensor can be decomposed into purely temporal and purely spatial parts with respect to this field. Let \(u^a\) be a timelike, normalized congruence of observers, then the projector onto the local space orthogonal to \(u^a\) is written as
\begin{equation}
h_{ab}\equiv g_{ab}+u_a u_b,
\qquad
h^a{}_b u^b=0.
\end{equation}

On the other hand, the covariant derivative of \(u^a\) admits the irreducible decomposition
\begin{equation}
\label{eq:kin_decomp_1p3}
\nabla_b u_a
= -A_a u_b
+\frac13\Theta\,h_{ab}
+\sigma_{ab}
+\omega_{ab},
\end{equation}
where $A_a \equiv u^b\nabla_b u_a$ is the acceleration, $\Theta \equiv \nabla_a u^a$ is the expansion, $\sigma_{ab} \equiv D_{\langle a}u_{b\rangle}$ is the shear (symmetric, trace-free, spatial), and $\omega_{ab} \equiv D_{[a}u_{b]}$ is the vorticity (antisymmetric, spatial). Here \(D_a\) denotes the projected spatial derivative, and angular brackets \(\langle\ \rangle\) indicate the symmetric, trace-free part orthogonal to \(u^a\).

Since any vector \(V^a\) can be decomposed into a temporal part plus a spatial part of the form
\begin{equation}
\label{descomposicion}
V^a=(-u_bV^b)u^a + h^a{}_b V^b,
\end{equation}
then, given a light ray with tangent vector \(k^a\), using the fact that the energy measured by \(u^a\) is \(E=-u_a k^a\), and that the spatial direction of propagation of the ray in the local frame of \(u^a\) is defined as
\begin{equation}
\label{eq:e_def}
e^a \equiv \frac{1}{E}\,h^a{}_b k^b,
\qquad u_a e^a=0,\qquad e_a e^a=1,
\end{equation}
and using \eqref{descomposicion}, one obtains the fundamental decomposition (e.g. \cite{TCM2008})
\begin{equation}
\label{eq:k_decomp_1p3}
k^a = E\left(u^a + e^a\right).
\end{equation}

The next step is to start from \(E=-u_a k^a\) and differentiate along the ray:
\begin{equation}
k^b\nabla_b E
= -\,k^b\nabla_b\left(u_a k^a\right)
= -\,k^a k^b\nabla_b u_a \;-\; u_a\,k^b\nabla_b k^a.
\end{equation}
If \(k^a\) satisfies the affine geodesic equation \(k^b\nabla_b k^a = 0\), the last term vanishes and one obtains
\begin{equation}
\label{eq:Eprop_basic}
k^b\nabla_b E = -\,k^a k^b\nabla_b u_a.
\end{equation}
Substituting \eqref{eq:kin_decomp_1p3} and the decomposition \eqref{eq:k_decomp_1p3}, one finds after direct contraction (e.g. \cite{EllisElst1998,Ellis2012})
\begin{equation}
\label{eq:dE_dlambda}
\frac{dE}{d\lambda}
\equiv k^a\nabla_a E
=
-\,E^2\left(
\frac13\Theta
+\sigma_{ab}e^a e^b
+A_a e^a
\right).
\end{equation}
The vorticity term \(\omega_{ab}\) does not contribute because \(\omega_{ab}\) is antisymmetric whereas \(e^a e^b\) is symmetric.

It is convenient to reparametrize the ray by introducing the \emph{optical parameter} \(\eta\), defined by
\begin{equation}
\label{eq:eta_def_1p3}
\frac{d}{d\eta}\equiv \frac{1}{E}\frac{d}{d\lambda}
\qquad\Longleftrightarrow\qquad
d\eta = E\,d\lambda.
\end{equation}
In this way, the propagation equation takes a simpler form for \(\ln E\). Thus, \eqref{eq:dE_dlambda} can be rewritten as
\begin{equation}
\label{eq:dlnE_deta}
\frac{d\ln E}{d\eta}
=
-\left(
\frac13\Theta
+\sigma_{ab}e^a e^b
+A_a e^a
\right),
\end{equation}

Integrating between emission and observation yields a covariant integral expression for the redshift:
\begin{equation}
\label{eq:z_integral_general}
\ln\!\left(\frac{E_e}{E_o}\right)
=
\int_{\eta_o}^{\eta_e}
\left(
\frac13\Theta
+\sigma_{ab}e^a e^b
+A_a e^a
\right)d\eta.
\end{equation}
This equation provides the ``transport mechanism'' that allows one to relate \(k^a\) at emission and observation. It gives us the ``covariant prescription'' for calculating the ratio \(\frac{E_e}{E_o}\) from what happens along the ray trajectory, and thus for evaluating Eq.~\eqref{eq:z_cov_def}.

\subsection{Standard result in FLRW}
\label{subsec:FLRW_from_1p3}

In an FLRW universe, and for the fundamental comoving congruence, we know that
\begin{equation}
A_a=0,\qquad \sigma_{ab}=0,\qquad \omega_{ab}=0,\qquad \Theta=3H
\end{equation}
where $H$ is the Hubble parameter. Then \eqref{eq:dlnE_deta} reduces to
\begin{equation}
\label{lnEFLRW}
\frac{d\ln E}{d\eta}=-H.
\end{equation}
Starting from Eq.~\eqref{lnEFLRW}, which is equivalent to
\begin{equation}
d\ln E=-H\,d\eta,
\end{equation}
and using \(H\equiv \dot a/a\), it is enough to relate \(d\eta\) to the cosmic time \(t\) along the ray. In standard comoving FLRW coordinates we take \(x^a=(t,x^i)\), where \(t\) is the \emph{cosmic time}. For a comoving observer (\(x^i=\) const.) one has \(d\tau=dt\), and therefore the 4-velocity is \(u^a=dx^a/d\tau=(1,0,0,0)\equiv(\partial_t)^a\). On the other hand, the tangent vector to the ray is \(k^a=dx^a/d\lambda\), so that its temporal component is \(k^t=dt/d\lambda\).
Hence, by contraction,
\[
E\equiv -u_a k^a = -u_t k^t = k^t = \frac{dt}{d\lambda}.
\]

On the other hand, the optical parameter \(\eta\) was defined by \(d\eta=E\,d\lambda\) [Eq. \eqref{eq:eta_def_1p3}], and therefore
\begin{equation}
d\eta=E\,d\lambda=dt,
\end{equation}
that is, along the ray and for comoving observers, \(d\eta\) coincides with \(dt\), up to the normalization associated with the choice of affine parameter, which cancels in ratios such as \(E_e/E_o\).
Therefore, Eq.~\eqref{lnEFLRW} can be rewritten as
\begin{equation}
\frac{d\ln E}{dt}=-H=-\frac{\dot a}{a}.
\end{equation}
Integrating,
\begin{equation}
d\ln E=-\frac{\dot a}{a}\,dt=-\,d\ln a
\qquad\Longrightarrow\qquad
\ln E=-\ln a+\text{const.},
\end{equation}
and finally
\begin{equation}
E\,a=\text{const.}
\qquad\Longrightarrow\qquad
E\propto \frac{1}{a}.
\end{equation}
Since \(1+z=E_e/E_o\), one immediately concludes that
\begin{equation}
\label{zflrw}
1+z_{\rm FLRW}=\frac{E_e}{E_o}=\frac{a_o}{a_e}.
\end{equation}

The symmetries present in the FLRW case allow this result to be obtained in several ways. Since in FLRW \(k^a\) has a known form, one may also use directly the relation between frequencies given by \eqref{eq:z_cov_def}, without going through \eqref{eq:dlnE_deta}.

\subsection{Redshift measured by tilted observers}
\label{subsec:tilted_master}

Let \(u^a\) be the comoving congruence of the background (the Hubble/CMB frame), and \(\tilde u^a\) the 4-velocity of a \emph{tilted} observer. The relation between them can be written as a Lorentz boost (e.g. \cite{TCM2008}):
\begin{equation}
\label{eq:tilt_u}
\tilde u^a=\tilde\gamma\left(u^a+\tilde v^a\right),
\qquad
u_a\tilde v^a=0,
\qquad
\tilde\gamma=(1-\tilde v^2)^{-1/2},
\qquad
\tilde v^2\equiv \tilde v_a\tilde v^a,
\end{equation}
where \(\tilde v^a\) is the peculiar (spatial) velocity of the tilted observer with respect to \(u^a\).
For non-relativistic flows, \(\tilde v^2\ll 1\), one has \(\tilde\gamma\simeq 1\) and \(\tilde u^a\simeq u^a+\tilde v^a\).

Let us now take \(s^a\) to be the unit vector (spatial with respect to \(u^a\)) pointing on the sky from the observer toward the source. We define the angle \(\alpha_s\) between \(\tilde v^a\) and \(s^a\) by
\begin{equation}
\label{eq:cosalpha_def}
\tilde v_a s^a \equiv \tilde v\,\cos\alpha_s,
\qquad
\tilde v\equiv \sqrt{\tilde v^2},
\qquad
\tilde v^2\equiv \tilde v_a\tilde v^a.
\end{equation}

Since \(E\equiv -u\!\cdot\!k\) and \(\tilde E\equiv -\tilde u\!\cdot\!k\), and using \(k^a=E(u^a+e^a)\) together with Eq.~\eqref{eq:tilt_u}, we can write
\begin{align}
\tilde E
&= -\,\tilde u_a k^a
= -\,\tilde\gamma(u_a+\tilde v_a)\,E(u^a+e^a) \nonumber\\
&= \tilde\gamma\,E\left[-u_a u^a - \tilde v_a e^a\right]
= \tilde\gamma\,E\left[1-\tilde v_a e^a\right].
\label{eq:Etilde_step}
\end{align}
For the direction observed in the sky \(s^a\), one has \(e^a=-s^a\). Therefore,
\begin{equation}
\label{eq:Etilde}
\tilde E = \tilde\gamma\,E\left(1+\tilde v_a s^a\right)
= \tilde\gamma\,E\left(1+\tilde v\cos\alpha_s\right).
\end{equation}
Motivated by this relation, we now introduce the directional relativistic \emph{Doppler factor}
\begin{equation}
\label{eq:D_def}
D \equiv \tilde\gamma\left(1+\tilde v\cos\alpha_s\right),
\qquad\Rightarrow\qquad
\tilde E = D\,E.
\end{equation}

Applying \eqref{eq:z_cov_def} to the energies measured by the real (tilted) observers, and using Eq.~\eqref{eq:D_def} at emission and observation, we obtain
\begin{equation}
\label{eq:master_zobs}
1+z_{\rm obs}
=\frac{\tilde E_e}{\tilde E_o}
=
\left(\frac{E_e}{E_o}\right)\frac{D_e}{D_o}.
\end{equation}
If the geometric background is FLRW, we have seen that the transport mechanism simplifies. In this case the non-local contributions reduce, so that \(E_e/E_o=a_o/a_e\), and therefore one obtains
\begin{equation}
\label{eq:master_zobs_FLRW}
1+z_{\rm obs}
=
\frac{a_o}{a_e}\;
\frac{D_e}{D_o}
=
\frac{a_o}{a_e}\;
\frac{\tilde\gamma_e\left(1+\tilde v_e\cos\alpha_{e,s}\right)}
{\tilde\gamma_o\left(1+\tilde v_o\cos\alpha_{o,s}\right)}.
\end{equation}
This is the equation that we shall use in observational applications, since it is written in terms of the observed direction on the sky \(s^a\) and local Doppler factors \(D\) evaluated at the emitter and at the observer.

Let us note that Eq.~\eqref{eq:master_zobs_FLRW} was obtained kinematically from the covariant definition of redshift \eqref{eq:z_cov_def} and from the local transformation of the measured energy, \(\tilde E = D\,E\), Eq.~\eqref{eq:D_def}. It is useful to clarify here how this result connects with the \(1+3\) propagation equation discussed above, where the energy \(E\) evolves along the ray according to Eq.~\eqref{eq:dE_dlambda}, or, in terms of the optical parameter \(\eta\), according to Eq.~\eqref{eq:dlnE_deta} and its integral form given by Eq.~\eqref{eq:z_integral_general}.

Since the null vector \(k^a\) characterizing the null ray is the same independently of the observer, the difference between measuring \(E=-u\!\cdot\!k\) or \(\tilde E=-\tilde u\!\cdot\!k\) lies only in the projection onto the observer's 4-velocity. In particular, from \(\tilde E = D\,E\) one immediately obtains the identity
\begin{equation}
\label{eq:lnE_split_D}
\frac{d\ln\tilde E}{d\eta}
=
\frac{d\ln E}{d\eta}
+
\frac{d\ln D}{d\eta},
\end{equation}
valid along the null geodesic parametrized by \(\eta\).
The last term, \(d\ln D/d\eta\), is the one that in the more general analysis appears as the derivative along the ray of \(\ln\!\big[\tilde\gamma(1-\tilde v_a e^a)\big]\) (equivalently, using \(e^a=-s^a\), of \(\ln\!\big[\tilde\gamma(1+\tilde v_a s^a)\big]\)).
This term encodes the variation of the directional Doppler factor when \(\tilde v^a\) (and/or the propagation direction) changes along the path of the ray, and it can be re-expressed in terms of gradients of the peculiar flow or, equivalently, in terms of the kinematical quantities of the tilted frame.

In this subsection, however, it is not necessary to integrate \eqref{eq:lnE_split_D} explicitly in order to obtain the observed \emph{instantaneous redshift}. Indeed, integrating between emission and observation one has \(\int_{\eta_o}^{\eta_e}(d\ln D/d\eta)\,d\eta=\ln(D_e/D_o)\), so that the whole contribution associated with \(d\ln D/d\eta\) reduces to the ratio of Doppler factors evaluated at the endpoints. It is enough to evaluate the Doppler factor \(D\) at the emission and observation events, which leads directly to Eq.~\eqref{eq:master_zobs} and, for an FLRW background, to the operational form shown in Eq.~\eqref{eq:master_zobs_FLRW}.
In the next section, when analyzing the \emph{redshift drift}, the time derivative of the ratio \(D_e/D_o\) (i.e. the evolution of \(\tilde v^a\) and of the angles \(\alpha_s\)) will play a central role; there the structure indicated by Eq.~\eqref{eq:lnE_split_D} becomes especially useful as a conceptual guide.

Now, let us show what the non-relativistic limit at first order in \(\tilde v\) looks like. Taking \(\tilde v\ll 1\), \(\tilde\gamma\simeq 1+\mathcal O(\tilde v^2)\), \(D\simeq 1+\tilde v\cos\alpha_s\), and using \(\frac{1}{1+x}\simeq(1-x)\), Eq.~\eqref{eq:master_zobs_FLRW} implies
\begin{equation}
\label{zobs_linear_pre}
1+z_{\rm obs}\simeq \frac{a_o}{a_e}\left(1+\tilde v_e\cos\alpha_{e,s}-\tilde v_o\cos\alpha_{o,s}\right),
\end{equation}
or equivalently, using \eqref{zflrw},
\begin{equation}
\label{eq:zobs_linear}
z_{\rm obs}\simeq z_{\rm FLRW} + (1+z_{\rm FLRW})\left(\tilde v_e\cos\alpha_{e,s}-\tilde v_o\cos\alpha_{o,s}\right).
\end{equation}

An important comment is in order regarding what changes, and what does not, when addressing the tilted case. Equation \eqref{eq:master_zobs_FLRW} shows that, at the kinematical level considered here, the effect of the \emph{tilt} is incorporated as a directional Doppler rescaling \(D_e/D_o\) multiplying the background redshift.
In particular, when studying the redshift drift, it will be convenient to keep the Hubble parameter \(H(z)\) as the background function associated with the FLRW geometry, and to compute \(dz_{\rm obs}/dt\) from Eq.~\eqref{eq:master_zobs_FLRW}. The corrections associated with the tilted frame enter through the time evolution of the scale factor \(a(t)\) and of the Doppler factors \(D\) (i.e. of \(\tilde v\) and of the angles \(\alpha_s\)).

A related recent analysis \cite{Pasten2026} has shown that a non-geodesic observational congruence can induce a dipolar contribution to the cosmological redshift through the line of sight projection of the observer 4-acceleration. This observation is consistent with the role of the acceleration term in the covariant \(1+3\) propagation equation above. Here we use the same covariant framework for a different purpose: starting from the general definition of redshift, we derive the full \(1+3\) propagation equation, recover the FLRW limit, and obtain for tilted observers Eq.~\eqref{eq:master_zobs_FLRW}, where the effect of the tilt appears as a ratio of Doppler factors evaluated at emission and observation. In this form, the Doppler contribution is explicitly reduced to the boundary factor \(D_e/D_o\), which will be the relevant quantity when deriving the redshift drift.

To conclude this subsection, let us make a few remarks that will be conceptually relevant for the next section. In many works addressing the \emph{tilted universe} approach, the ``apparent acceleration'' associated with a bulk flow is discussed in terms of the deceleration parameter measured by the tilted congruence, \(\tilde q\). This quantity is a well-defined local kinematical scalar associated with the congruence of tilted observers. However, its connection with cosmological data is not completely direct, since quantities such as \(q\) or \(\tilde q\) are defined locally from the kinematics of a family of observers, whereas astronomical observables are constructed from light signals received on the past light cone. For this reason, observables such as the redshift and the redshift drift are more directly tied to what is measured, since they are defined through the relation between emitter, light ray, and observer. By contrast, connecting a local parameter such as \(\tilde q\) with an observational cosmographic inference requires additional interpretive steps. This motivates the use of light-cone observables when assessing the physical relevance of the \emph{tilted} scenario.

The use of \emph{effective} parametrizations remains useful, for example, when fitting the supernova Hubble diagram through luminosity distances, which are integrated light-cone quantities. This is the strategy followed in \cite{Asvesta2022}, where the effect of the tilt is encoded in a parametrization of \(\tilde q(z)\). From this function, an effective expansion history \(\tilde H_{\rm eff}(z)\) is reconstructed and the luminosity distance \(d_L(z)\) is then computed and fitted to the Pantheon sample. In that sense, a prediction based entirely on such an effective function may be driven close to the accelerated \(\Lambda\)CDM behavior by construction, since the effective expansion history is adjusted precisely to reproduce the observed Hubble diagram. Within this phenomenological approach, the \emph{tilted universe} scenario shows that the acceleration inferred from supernova data could, in principle, be interpreted as an apparent effect associated with peculiar motions rather than as evidence for a cosmological constant. However, for tests such as the redshift drift, this procedure may obscure the separation between the background expansion and the local kinematical corrections, because the relevant peculiar-motion effects are packaged into an effective function such as \(\tilde q(z)\) or \(\tilde H_{\rm eff}(z)\). Therefore, in what follows we adopt a complementary strategy: we specify a concrete background expansion history \(H(z)\), such as Einstein--de Sitter or \(\Lambda\)CDM, and compute on top of it the corrections induced by the relative motion of the tilted observers. This allows us to distinguish more explicitly between the contribution of the assumed background spacetime and the contribution of local tilted kinematics to the predicted redshift drift.

This methodological choice is also especially pertinent in light of \cite{ClarksonMaartens2026,Tsagas2026reply,PatliakaTsagas26}. Even if the direct interpretation of \(\tilde q\) as an observed cosmographic parameter and the underlying peculiar dynamics remain under debate today, the treatment of the redshift drift in terms of the actually measured redshift, the adopted FLRW background, and the Doppler factors evaluated at emission and observation remains of clear interest as a well-posed observational problem.

In what follows, the sign of the redshift drift will be used only as a diagnostic of the specific model under consideration. In a strictly FLRW spacetime, this sign is directly related, at low redshift, to the background deceleration parameter. In the tilted case, by contrast, the drift contains an additional direction-dependent peculiar motion contribution, derived explicitly in Sec.~III.A. Hence, its sign cannot be interpreted as a model-independent indicator of global acceleration of the background universe.

A related light-cone construction is the null deceleration parameter discussed in Sec.~7.1.10 of Ref.~\cite{Tsagas2025rev}, where the effect of peculiar motion is analysed for cosmographic quantities associated with null congruences. This is conceptually close to the present discussion, since both approaches emphasize that relative motion affects quantities inferred from light propagation. However, our observable is not the null deceleration parameter itself, but the redshift drift \(dz_{\rm obs}/d\tau_o\), obtained directly from the covariant redshift relation and its time variation. For this reason, we keep the drift correction explicitly in the form \(\delta\dot z(\hat s,z)\), rather than rewriting it in terms of an effective deceleration parameter.

\section{Redshift drift in the tilted scenario}
\label{sec_3}

In this section we derive the \emph{redshift drift} signal measured by a tilted observer at first order in the relative motion. Starting from the Doppler-rescaled redshift relation, we separate the standard FLRW Sandage--Loeb contribution from a local kinematical correction sourced by the peculiar expansion, shear, and acceleration of the tilted congruence. We then introduce a minimal dipolar approximation motivated by the observed large-scale \emph{bulk flow} direction, parametrizing the loss of coherence with redshift through a transition scale. Finally, we use this phenomenological model to estimate the expected signal in both an Einstein--de Sitter background and the standard \(\Lambda\)CDM case.

\subsection{Exact first-order solution}
\label{firstordersol}

The redshift drift detected by a \emph{tilted} observer is defined as
\begin{equation}\label{def_reddrift}
\zdot \equiv \frac{d\zobs}{d\tau_o},
\end{equation}
where \(\tau_o\) is the proper time of the observer at the observation event. Since we shall assume a small tilt, \(\tilde v^2\ll 1\), one has \(\tilde\gamma\simeq 1\), and therefore, to first order, the proper times satisfy \(d\tau_o\simeq d\tilde\tau_o\). The key subtlety is that, when going from \(\tau_o\) to \(\tau_o+d\tau_o\), the emission event also changes. Therefore, when differentiating any quantity \(X\) evaluated at emission, we have
\begin{equation}
\frac{d}{d\tau_o}X_e=\frac{d\tau_e}{d\tau_o}\,\frac{d}{d\tau_e}X_e.
\end{equation}
On the other hand, in general, and in a fully covariant way, the identity
\begin{equation}\label{eq:taurelation}
\frac{d\tau_e}{d\tau_o}=\frac{(k_a\tilde u^a)_o}{(k_a\tilde u^a)_e}=\frac{1}{1+\zobs}
\end{equation}
holds. This is a reformulation of the fact that the number of wave crests between two consecutive events at the observer and at the emitter is the same.

Now, if the geometric background is FLRW, we have seen that, according to Eq.~\eqref{eq:master_zobs_FLRW}, one can write
\begin{equation}
\label{eq:master_obs2}
1+z_{\rm obs}
=
\frac{a_o}{a_e}\;
\frac{D_e}{D_o}.
\end{equation}

Starting from this last equation and taking the logarithm, we obtain
\begin{equation}\label{eq:logdiff1}
\ln(1+\zobs)=\ln a_o-\ln a_e+\ln D_e-\ln D_o.
\end{equation}
Differentiating with respect to \(\tau_o\), we find
\begin{equation}\label{eq:logdiff2}
\frac{\zdot}{1+\zobs}=
\frac{d}{d\tau_o}\ln a_o-\frac{d}{d\tau_o}\ln a_e
+\frac{d}{d\tau_o}\ln D_e-\frac{d}{d\tau_o}\ln D_o.
\end{equation}

Since \(d(\ln a)/d\tau=\frac{1}{a}\frac{da}{d\tau}\equiv H\) in the comoving frame (and \(d\tau\simeq d\tilde\tau\) to first order), we have
\begin{equation}
\frac{d}{d\tau_o}\ln a_o=H_o.
\end{equation}
For the emission term we use \eqref{eq:taurelation}:
\begin{equation}
\frac{d}{d\tau_o}\ln a_e
=\frac{d\tau_e}{d\tau_o}\,\frac{d}{d\tau_e}\ln a_e
=\frac{1}{1+\zobs}\,H_e.
\end{equation}
Substituting into \eqref{eq:logdiff2} and, for the moment, ignoring the Doppler contribution, we recover the standard term:
\begin{equation}\label{eq:SLtilted}
\left.\frac{\zdot}{1+\zobs}\right|_{\rm FLRW}
=H_o-\frac{H_e}{1+\zobs}
\quad\Longrightarrow\quad
\zdot_{\rm FLRW}=(1+\zobs)H_o-H_e.
\end{equation}
This is the background Sandage--Loeb term evaluated at the observed redshift \cite{Sandage1962,Loeb1998}.

We now define
\begin{equation}\label{eq:deltazdotdef}
\delta\zdot \equiv (1+\zobs)\left(\frac{d}{d\tau_o}\ln D_e-\frac{d}{d\tau_o}\ln D_o\right).
\end{equation}
With this definition, \eqref{eq:logdiff2} can be rewritten exactly as
\begin{equation}\label{eq:split}
\zdot=\underbrace{\bigl[(1+\zobs)H_o-H_e\bigr]}_{\zdot_{\rm FLRW}}+\delta\zdot.
\end{equation}

It is useful to recall how the first term in Eq.~\eqref{eq:split} is related to the standard deceleration parameter in the strictly FLRW limit. Setting the peculiar contribution to zero, \(\delta\dot z=0\), and writing \(z_{\rm obs}=z\), one has
\[
\dot z_{\rm FLRW}=(1+z)H_0-H(z).
\]
At low redshift, the background Hubble parameter can be expanded as
\[
H(z)=H_0\left[1+(1+q_0)z+\mathcal{O}(z^2)\right],
\]
where
\[
q_0\equiv -\frac{\ddot a_0}{a_0H_0^2}
\]
is the present FLRW deceleration parameter. Therefore,
\[
\dot z_{\rm FLRW}
=
(1+z)H_0-H(z)
=
-q_0H_0z+\mathcal{O}(z^2).
\]
Thus, in a strictly FLRW spacetime, the sign of the (low-\(z\)) redshift drift is directly related to the sign of the background deceleration parameter \(q_0\). This statement, however, refers only to the FLRW contribution. For tilted observers, the observed drift also contains the peculiar-motion correction defined later in Eq.~\eqref{deltazdot}; therefore, the FLRW connection between the sign of the drift and the background deceleration parameter cannot be applied to the full observed drift without specifying the tilted contribution.

Let us note that, if the physical background is taken to be EdS, for example, it is enough to use \(H_e=H_{\rm EdS}(z_{\rm obs})=H_0(1+z_{\rm obs})^{3/2}\) and \(H_o=H_0\) in Eq.~\eqref{eq:split}, in order to obtain:
\begin{equation}
\zdot=\zdot_{\rm EdS}+\delta\zdot,
\qquad
\zdot_{\rm EdS}=(1+\zobs)H_0-H_0(1+\zobs)^{3/2}.
\end{equation}
We will return to these expressions at the end, when presenting and analyzing the results.

The next step is to use the chain rule again for the emitter term, and to rewrite Eq.~\eqref{eq:deltazdotdef} as
\begin{equation}\label{eq:deltazdot_expand}
\delta\zdot
=(1+\zobs)\left[\frac{1}{1+\zobs}\frac{d}{d\tau_e}\ln D_e-\frac{d}{d\tau_o}\ln D_o\right]
= \frac{d}{d\tau_e}\ln D_e-(1+\zobs)\frac{d}{d\tau_o}\ln D_o.
\end{equation}

Based on what we have seen in subsections \ref{redshift_cov} and \ref{subsec:tilted_master}, in the \(1+3\) formalism and for the congruence of tilted observers \(\tilde u^a\), we perform the kinematical decomposition
\begin{equation}\label{gradtilde_u}
\nabla_b \tilde u_a
=
\frac13\tilde\Theta\,\tilde h_{ab}
+\tilde\sigma_{ab}
+\tilde\omega_{ab}
-\tilde A_a \tilde u_b,
\end{equation}
where \(\tilde h_{ab}=g_{ab}+\tilde u_a\tilde u_b\) is the spatial projector. For a light ray with null wave vector \(k^a\), the energy measured by the tilted observer is \(\tilde E\equiv -k_a\tilde u^a\), and the spatial direction measured by that observer is \(\tilde e^a\), defined through the decomposition $k^a=\tilde E\bigl(\tilde u^a+\tilde e^a\bigr)$, with \(\tilde u_a\tilde e^a=0\) and \(\tilde e_a\tilde e^a=1\).

Using the geodesic equation for the ray, \(k^b\nabla_b k^a=0\), together with the decomposition in Eq.~\eqref{gradtilde_u}, one obtains the standard propagation equation for the measured energy:
\begin{equation}\label{eq:propE}
\frac{1}{\tilde E}\frac{d\tilde E}{d\tau}
=
-\left(
\frac13\tilde\Theta
+\tilde\sigma_{ab}\tilde e^a\tilde e^b
+\tilde A_a\tilde e^a
\right),
\end{equation}
where, for simplicity, we have rewritten the affine parameter \(\lambda\) through $\frac{d}{d\lambda}=\tilde E\,\frac{d}{d\tau}$.

On the other hand, we had seen in Eq.~\eqref{eq:D_def} that \(\tilde E=D\,E\), so that \(\ln D=\ln\tilde E-\ln E\). Differentiating, we can therefore write
\begin{equation} \label{dlnD}
\frac{d}{d\tau}\ln D
=
\frac{1}{\tilde E}\frac{d\tilde E}{d\tau}
-
\frac{1}{E}\frac{dE}{d\tau}.
\end{equation}
From this relation, we can already see that the derivatives of the \(\ln D\) terms appearing in Eq.~\eqref{eq:deltazdot_expand} will be obtained from the energy propagation equations, both for the background and for the tilted frame.

Starting from Eq.~\eqref{eq:propE}, and using the fact that the direction toward the source on the sky is \(\tilde e^a=-\hat s^a\), we now define the tilted ``directional Hubble'' as
\begin{equation} \label{HubbleDir}
\tilde{\mathcal H}(\hat s)
=
\frac{1}{3}\tilde\Theta
+\tilde\sigma_{ab}\hat s^a\hat s^b
-\tilde A_a\hat s^a.
\end{equation}
Then,
\begin{equation} \label{Hache}
\frac{1}{\tilde E}\frac{d\tilde E}{d\tau}
=
-\tilde{\mathcal H}(\hat s).
\end{equation}

On the other hand, with respect to the comoving background \(u^a\), and using the fact that in the FLRW case \(A_a=0\), \(\sigma_{ab}=0\), and \(\Theta=3H\), we obtain
\begin{equation} \label{HacheFLRW}
\frac{1}{E}\frac{dE}{d\tau}
=
-H.
\end{equation}

In the tilted scenario, to first order in \(\tilde v^a\), one introduces the peculiar volume scalar \(\tilde\vartheta\), defined by
\begin{equation}
\tilde\Theta = \Theta + \tilde\vartheta,
\label{eq:expansion-split}
\end{equation}
which follows from \(\tilde\Theta=\nabla_a\tilde u^a\), together with $\nabla_a\tilde v^a\simeq D_a\tilde v^a\simeq \tilde D_a\tilde v^a$, e.g. \cite{Tsagas2013}. Here, \(\tilde\vartheta\equiv \tilde D_a\tilde v^a\), and it measures the average expansion or contraction of the peculiar flow with respect to the background. When \(\tilde\vartheta<0\), the local flow is contracting on average, corresponding to a \emph{contractive bulk flow}, whereas \(\tilde\vartheta>0\) corresponds to a relative expansion. Recall that in the tilted frame the spatial derivative is \(\tilde D_a=\tilde h_a{}^b\nabla_b\), with \(\tilde h_{ab}\simeq h_{ab}\) at the order considered here.

Using \eqref{eq:expansion-split}, we define the purely \emph{peculiar tilted correction} on the spacetime event $x$ as
\begin{equation}\label{Hpec}
\mathcal H_{\rm pec}(\hat s; x)
\equiv
\tilde{\mathcal H}(\hat s;x)-H(x)
=
\frac{1}{3}\tilde\vartheta(x)
+\tilde\sigma_{ab}(x)\hat s^a\hat s^b
-\tilde A_a(x)\hat s^a.
\end{equation}
In the last equation, the first term involving \(\tilde\vartheta\) is the isotropic part of the peculiar expansion; the second term, containing the shear \(\tilde\sigma_{ab}\), describes the angular deformation or anisotropy of the flow; and the last term gives the peculiar acceleration projected along the line of sight.

With these definitions, Eq.~\eqref{dlnD} simply becomes
\begin{equation}\label{dlnDHpec}
\frac{d}{d\tau}\ln D
=
-\mathcal H_{\rm pec}(\hat s;x)
\end{equation}
and therefore we have
\begin{equation}
\frac{d}{d\tau_e}\ln D_e
=
-\mathcal H_{{\rm pec},e}(\hat s,z),
\qquad
\frac{d}{d\tau_o}\ln D_o
=
-\mathcal H_{{\rm pec},o}(\hat s),
\end{equation}
where the subscripts \(e\) and \(o\) in \(\mathcal H_{\rm pec}\) indicate that the local quantities defined in Eq.~\eqref{Hpec} are evaluated at the emission and observation events, respectively. For this same reason, note that \(\mathcal H_{{\rm pec},e}\) is a function of both $\hat s$ and $z$.

Collecting all these results, and since in the background functions (at the first order at which we are working) we can replace \(\zobs\simeq z_{\rm FLRW}=z\), the redshift drift detected by a tilted observer, by virtue of Eq.~\eqref{eq:split} and Eq.~\eqref{eq:deltazdot_expand}, results
\begin{equation}\label{zdotcompacto}
\dot z = (1+z)H_0-H(z) + \delta\dot z(\shat,z)=\zdot_{\rm FLRW}(z)+\dzdot(\shat,z),
\end{equation}
where
\begin{equation}\label{deltazdot}
\delta\dot z(\hat s,z)
=
(1+z)\,\mathcal H_{{\rm pec},o}(\hat s)
-
\mathcal H_{{\rm pec},e}(\hat s,z).
\end{equation}

For tilted observers, a positive value of the observed redshift drift does not by itself imply global acceleration of the background universe, since the drift also contains the directional peculiar motion correction \(\delta\dot z(\hat s,z)\). Therefore, its sign can be interpreted only within a given background cosmology and a definite prescription for the tilted contribution.

\subsection{Dipolar approximation}
\label{dipolarsec}

In this subsection, we aim to estimate the expected value of the redshift drift signal that would be detected by a tilted observer. To achieve this, we shall introduce a set of physically motivated and observationally guided assumptions about certain quantities involved in the calculation presented in the previous subsection, as well as about their numerical order of magnitude. We will then present the results obtained within these approximations and perform the expected comparisons between two different theoretical frameworks, namely the EdS case and the \(\Lambda\)CDM model.

Equation \eqref{deltazdot} is sourced by gradients of the peculiar-velocity field, which have not yet been measured with sufficient precision. The full gradients, namely divergence and shear, require three-dimensional reconstructions or estimates that are not yet available to us. The work \cite{Pasten2024} provides, to the best of our knowledge, the most direct observational evidence that the local peculiar-velocity field may exhibit a negative mean divergence on certain scales, and that its spatial structure is not isotropic. However, that work does not yet reconstruct all the kinematical ingredients required to feed a complete observational prediction for the \emph{tilted redshift drift}, in particular its time dependence, the projected acceleration term, the emitter contribution, and the full angular modulation of the effect. Therefore, in what follows we adopt a minimal approximation that will allow us to estimate the order of magnitude of the expected redshift drift signal for a tilted observer, using the data analysed and reported from the CosmicFlows-4 (CF4) catalog for the detected \emph{bulk flow} \cite{Watkins2023}, also recently discussed in \cite{Watkins2025}. In this way, we connect the tilted correction \(\delta\dot z\) directly to a kinematical observable.

The data presented in \cite{Watkins2023} highlight that the large-scale flow is dominated by external contributions on scales of order \(\sim 200\,h^{-1}\,\mathrm{Mpc}\). For a sphere of radius \(R=200\,h^{-1}\,{\rm Mpc}\), a bulk flow with velocity \(U\) and direction $\hat e_{\rm BF}$ is reported as
\begin{equation}
\label{veloc_direc}
|U|=419\pm 36~{\rm km/s},
\qquad
(l,b)=(298^\circ\pm 5^\circ,\,-8^\circ\pm 4^\circ).
\end{equation}
This fixes the direction of the differential test, namely toward the bulk flow and toward its antipode, as well as the scale \(R\) of the coherent flow. A purely uniform bulk flow does not produce gradients. For this reason, we need an operational model: we take the bulk flow measured within a sphere of radius \(R\) as a coherent velocity variation across the volume. The simplest estimate of a characteristic gradient aligned with the bulk flow is
\begin{equation}
\label{eq:deltaH_def}
\delta H \equiv \frac{|U|}{R}.
\end{equation}
Here, \(\delta H\) constitutes a minimal phenomenological ansatz. It will play the role of the observational ``effective amplitude'', fixing the scale of the projected kinematical corrections, or equivalently the order of magnitude of the effective deformation rate of the peculiar field along the bulk flow direction. Setting \(R=200\,h^{-1}\,{\rm Mpc}\), adopting \(H_0=67.4\ {\rm km\,s^{-1}\,Mpc^{-1}}\) \cite{Planck:2018CosmologicalParameters}, and taking \(|U|=419\,{\rm km/s}\), one obtains
\[
\delta H=1.4120\ {\rm km\,s^{-1}\,Mpc^{-1}}.
\]

In the derivation based on the covariant \(1+3\) formalism, at linear order in the tilt, we obtained the redshift drift prediction
\[
\zdot(\shat,z)=\zdot_{\rm FLRW}(z)+\dzdot(\shat,z),
\]
with \(\dzdot(\shat,z)\) given by Eq.~\eqref{deltazdot}. There, \(\mathcal H_{\rm pec}(\hat s,z)\) compactly represents the line-of-sight projected contributions that appear in the covariant formalism, such as divergences, peculiar acceleration, shear, and related kinematical terms.

For a preferred direction \(\hat e=\hat e_{\rm BF}\), given by the bulk flow axis, we write the directional prediction as an angular expansion:
\begin{equation}
\dot z(\hat s,z)=\dot z_0(z)+\dot z_1(z)\cos\alpha+\text{\dots}\,,
\qquad
\text{where:}\quad \cos\alpha=\hat s \cdot \hat e .
\label{eq:zdot_multipole_def}
\end{equation}
Here \(\dot z_0\) is the monopole, namely the isotropic part, and \(\dot z_1\) is the dipole amplitude: it quantifies how much \(\dot z\) differs when looking along (\(\cos\alpha=+1\)) or against (\(\cos\alpha=-1\)) the flow. The dipole is, by definition, the coefficient of the term proportional to \(\cos\alpha\) in \(\dot z\).

In a model dominated by a single axis \(\hat e\), the local term at the observer defines a ``today'' dipole amplitude, which we write as
\begin{equation}
\label{Hpecobs}
\mathcal{H}_{{\rm pec},o}(\hat s)\equiv A_o\,\cos\alpha
\quad \Rightarrow\quad
(1+z)\,\mathcal{H}_{{\rm pec},o}=(1+z)A_o\cos\alpha,
\end{equation}
whereas the emitter term defines an amplitude at depth \(z\),
\begin{equation}
\label{Hpecemis}
\mathcal{H}_{{\rm pec},e}(\hat s,z)\equiv A_e(z)\,\cos\alpha.
\end{equation}
Thus, one immediately obtains
\begin{equation}
\dot z_1(z)= (1+z)\,A_o- A_e(z).
\label{eq:zdot_dipole_Ao_Ae}
\end{equation}

The exact \(1+3\) equation for \(\delta\dot z(\hat s,z)\) is not, in general, purely dipolar. When \(\mathcal H_{\rm pec}(\hat s)\) is expanded in the observation direction, there appear, in principle, a monopole contribution proportional to \(\tilde\vartheta\), a dipole contribution associated with vectorial projections along the line of sight, and a quadrupole contribution associated with the shear term \(\tilde\sigma_{ab}\hat s^a\hat s^b\). Therefore, the dipolar form adopted here should not be understood as an exact identity of the theory, but rather as a \emph{minimal multipolar truncation}. Its motivation is twofold. On the one hand, from a theoretical point of view, if the anisotropy is dominated by a coherent large-scale flow with a single preferred axis \(\hat e\), the first natural anisotropic multipole is the dipole, since any vectorial projection along the line of sight produces an angular dependence \(\hat s \cdot \hat e=\cos\alpha\), whereas shear and higher multipoles can be treated as subdominant corrections. On the other hand, beyond the theoretical motivation based on the multipolar structure of the exact equation, the choice of a minimal dipolar parametrization is also phenomenologically supported. Several observational analyses find signals compatible with the existence of a preferred direction in the local kinematics, or with coherent large-scale bulk flows, both in Type Ia supernovae within the tilted framework \cite{Sah2024} and in bulk flow reconstructions. In particular, the CosmicFlows-4 analyses of \cite{Watkins2023,Watkins2025} show the presence of a coherent large-scale bulk flow and study its physical origin. Since a bulk flow corresponds precisely to the dipolar component of the peculiar velocity field, these results reinforce the idea that, at least as a first effective model, the dominant anisotropy can be captured by retaining the dipole term, while shear and higher multipoles are treated as subdominant corrections. In this sense, the observational results do not prove that the exact redshift drift equation is purely dipolar, but they do support a dipolar truncation as a physically reasonable and phenomenologically well-motivated approximation for describing the leading anisotropy. Along these lines, the approximation
\[
\mathcal H_{\rm pec}(\hat s)\simeq A\,\cos\alpha
\]
should be interpreted as the dominant anisotropic term of a more general angular expansion, whose ultimate validity must be assessed a posteriori through comparison with the data.

In the tilted scenario, the dipolar contribution to the redshift drift should switch off as one observes more distant sources, beyond the coherent domain of the flow characterized by a scale \(L\sim\lambda_T\), which we may identify with the \emph{transition scale} \(\lambda_T\) typically discussed in the \emph{tilted universe} scenario, e.g. \cite{Tsagas2022}. We now introduce the ``coherence function'' \(F\), which models how much of the emitter contribution remains correlated with our local coherent domain when the source is at comoving distance \(\chi(z)\). The minimal parametrization we shall use is \(A_e(z)=A_o\,F(z;L)\), meaning that
\begin{equation}
F(z;L)\ \equiv\ \frac{A_e(z)}{A_o}.
\label{eq:F_unambiguous_AeAo}
\end{equation}
That is, \(F\) is exactly the ratio between the dipole amplitude at the emitter and the local dipole amplitude at the observer.

Then, starting from Eq.~\eqref{deltazdot} and introducing the definitions in Eqs.~\eqref{Hpecobs} and \eqref{Hpecemis}, we obtain
\begin{equation}
\dot z(\hat s,z)=\dot z_{\rm FLRW}(z)\;+\;\Big[(1+z)A_o-A_e(z)\Big]\cos\alpha.
\label{eq:zdot_line_by_line2}
\end{equation}
Comparing with Eq.~\eqref{eq:zdot_multipole_def} and using Eq.~\eqref{eq:F_unambiguous_AeAo}, we conclude that
\begin{equation}
\dot z_1(z)= (1+z)A_o-A_e(z)=A_o\Big[(1+z)-F(z;L)\Big].
\end{equation}
Let us note that the expected decrease of the dipole with redshift is encoded in the fact that \(F(z;L)\) should decrease once \(\chi(z)\) exceeds the coherence scale \(L=\lambda_T\).

We now propose
\begin{equation}
F(z;z_\star)\equiv e^{-z/z_\star},
\qquad \text{with}\quad \lambda_T\equiv \chi(z_\star),
\label{eq:F_our_choice}
\end{equation}
so that the damping with depth is controlled by \(z_\star\), or equivalently by \(L=\lambda_T\). In works related to the \emph{tilted universe} scenario, \(\lambda_T\) is a spatial transition scale, typically expressed in Mpc, which effectively marks the distance range over which the tilted observer may be dominated by kinematical terms of the peculiar-velocity field, such as combinations involving \(\tilde\vartheta/\Theta\) and/or projected anisotropic terms. Beyond that scale, the behavior tends to approach that of the FLRW background. Here, \(L=\lambda_T\) enters only through the distance dependence of \(\mathcal{H}_{{\rm pec},e}\), namely as a \emph{radial window} that suppresses the contribution of peculiar gradients once the source lies beyond the relevant coherence scale.

The next step is to identify the amplitude \(A_o\) with the observational quantity denoted by \(\delta H\), which finally leads to our final expression:
\begin{equation}
\label{deltazdotfinal}
\delta\dot z(\hat s,z)=\delta H\big[(1+z)-e^{-z/z_\star}\big]\cos\alpha.
\end{equation}
Let us note that, for \(z\ll 1\), a Taylor expansion of the exponential function gives \(\delta\dot z\propto z\), which implies that at \(z=0\) one obtains \(\delta\dot z=0\), as required.

\subsection{Results}
\label{Results}

In this subsection, we present the expected predictions for the redshift drift signal in the \emph{tilted} framework. We show our results for two different scenarios: first, for a matter-only universe (Einstein--de Sitter, EdS), and then for the standard \(\Lambda\)CDM case.

The logic of this comparison is the following. The central question addressed here is what imprint the observed large-scale \emph{bulk flows} would leave on the redshift drift if they are modeled as an effective dipolar contribution within a two-congruence relativistic framework. The Einstein--de Sitter case is used as the natural benchmark of the original tilted scenario and also as a theoretical control model. Since it contains no dark energy and is globally decelerating, any departure from the standard EdS drift can be attributed to the kinematical effect of the tilt rather than to a cosmological constant component. The complementary \(\Lambda\)CDM case is included not because the tilted correction is needed to produce acceleration in that background, but in order to quantify the possible directional bias induced by peculiar motion effects on top of the concordance model. In this way, keeping the background expansion \(H(z)\) explicit (instead of absorbing the effect into an effective \(\tilde H_{\rm eff}(z)\) or \(\tilde q(z)\) fitted to the Hubble diagram) makes the test less degenerate and allows a dipolar contribution to \(\dot z\) to be associated directly with the change of observer frame and with the phenomenological modeling of the local bulk flow.

For both scenarios we use the final expression obtained in the previous subsection,
\[
\zdot(\shat,z)=(1+z)\Hzero-\Hub(z)+\;\dzdot(\shat,z),
\]
where the tilted correction, approximated by a dipolar contribution, is given by Eq. \eqref{deltazdotfinal}.

Recall that here we will fix \(\delta H=1.4120\ {\rm km\,s^{-1}\,Mpc^{-1}}\), as discussed below the Eq. \eqref{eq:deltaH_def}. The value adopted for \(z_\star\) will be \(z_\star=0.1\), chosen simply because it lies within the range of values mentioned in \cite{Sah2024} from supernovae, where the authors find a dipolar signal in \(H(z)\). On the other hand, since we have taken \(\lambda_T=\chi(z_\star)\), the value \(z_\star=0.1\) implies \(\lambda_T\sim 400\,{\rm Mpc}\), both for the EdS and \(\Lambda\)CDM cases, in agreement with the typical values reported for local bulk flows; see, for instance, Table~2 in \cite{Tzartinoglou2024}.

The Hubble functions will be chosen as follows. For ``Test A'' (EdS), where \(\Omega_m=1\), we use
\[
H(z)=H_0(1+z)^{3/2}.
\]
For ``Test B'' (\(\Lambda\)CDM), adopting \(\Omega_m=0.315\) and \(\Omega_{\Lambda}=0.685\) from \cite{Planck:2018CosmologicalParameters}, we use
\[
H(z)=H_0\left[\Omega_m(1+z)^3+\Omega_{\Lambda}\right]^{1/2}.
\]
In both tests we take \(H_0=67.4\ {\rm km\,s^{-1}\,Mpc^{-1}}\).

Let us note that the tilted correction \(\delta\dot z\) is the same in both tests. What changes in each test is the function \(H(z)\) that is used. Regarding the direction of the observed bulk flow, the CF4 result in Eq.~\eqref{veloc_direc} provides a reference axis that may be identified, for definiteness, with
\[
(l,b)_{\rm BF}=(298^\circ,-8^\circ).
\]
The specific numerical direction is not essential for the results shown below, since we only consider the two extremal projections along and against the bulk flow axis, corresponding to \(\cos\alpha=\pm1\). Since \(\cos\alpha=\hat s \cdot \hat e_{{\rm BF}}\), for an arbitrary source with Galactic coordinates \((l_s,b_s)\), the projection onto this axis would be given by
\[
\cos\alpha
=
\sin b_s\,\sin b_{\rm BF}
+\cos b_s\,\cos b_{\rm BF}\,\cos\left(l_s-l_{\rm BF}\right).
\]
However, since here we are interested only in estimating the expected order of magnitude of the redshift drift signal, it is sufficient to consider the maximal projections,
\[
\cos\alpha=\pm1.
\]

\begin{figure}[htbp]
    \centering

    \begin{subfigure}{0.48\textwidth}
        \centering
        \includegraphics[width=\textwidth]{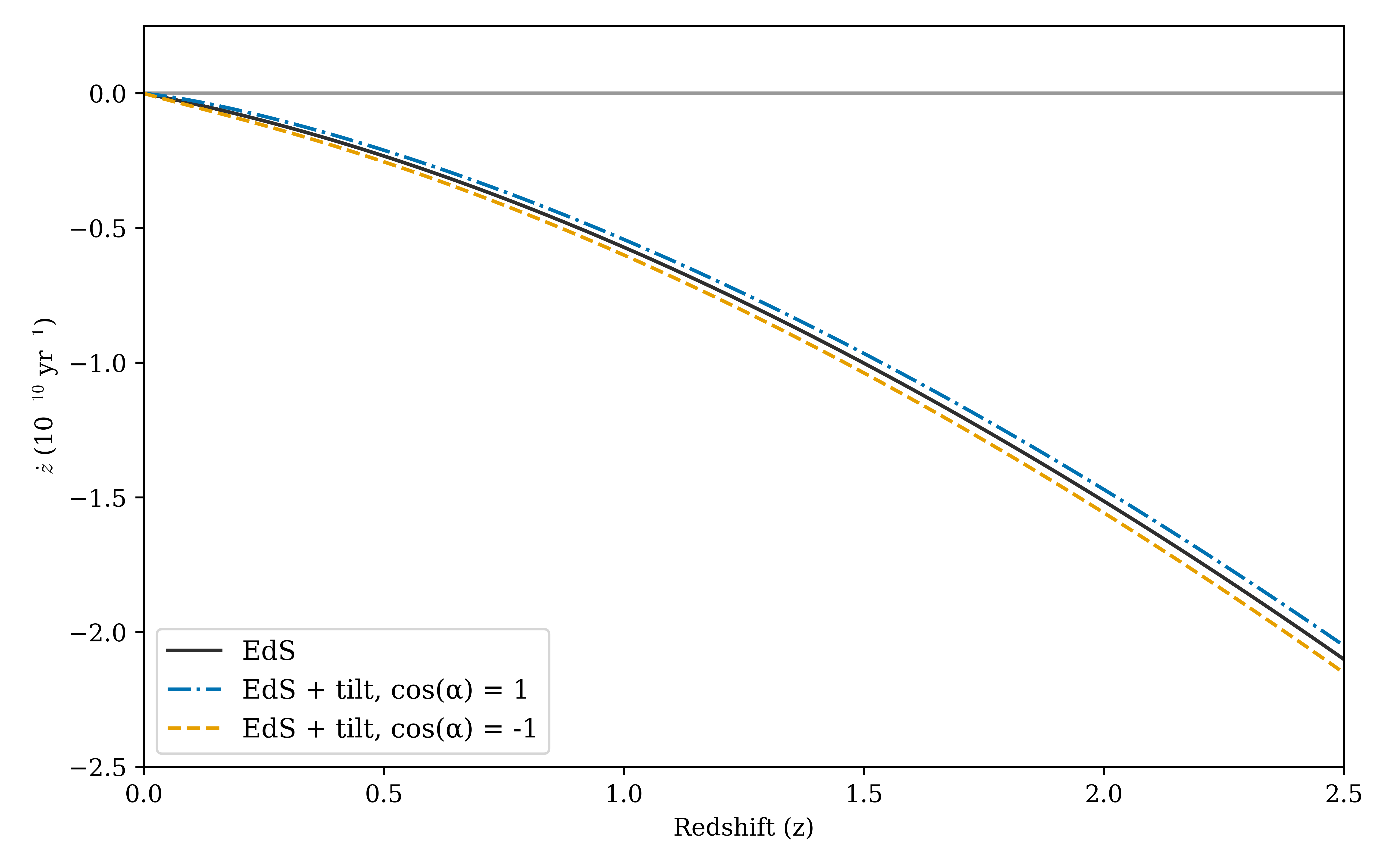}
        \caption{}
        \label{fig:EdS_full}
    \end{subfigure}
    \hfill
    \begin{subfigure}{0.48\textwidth}
        \centering
        \includegraphics[width=\textwidth]{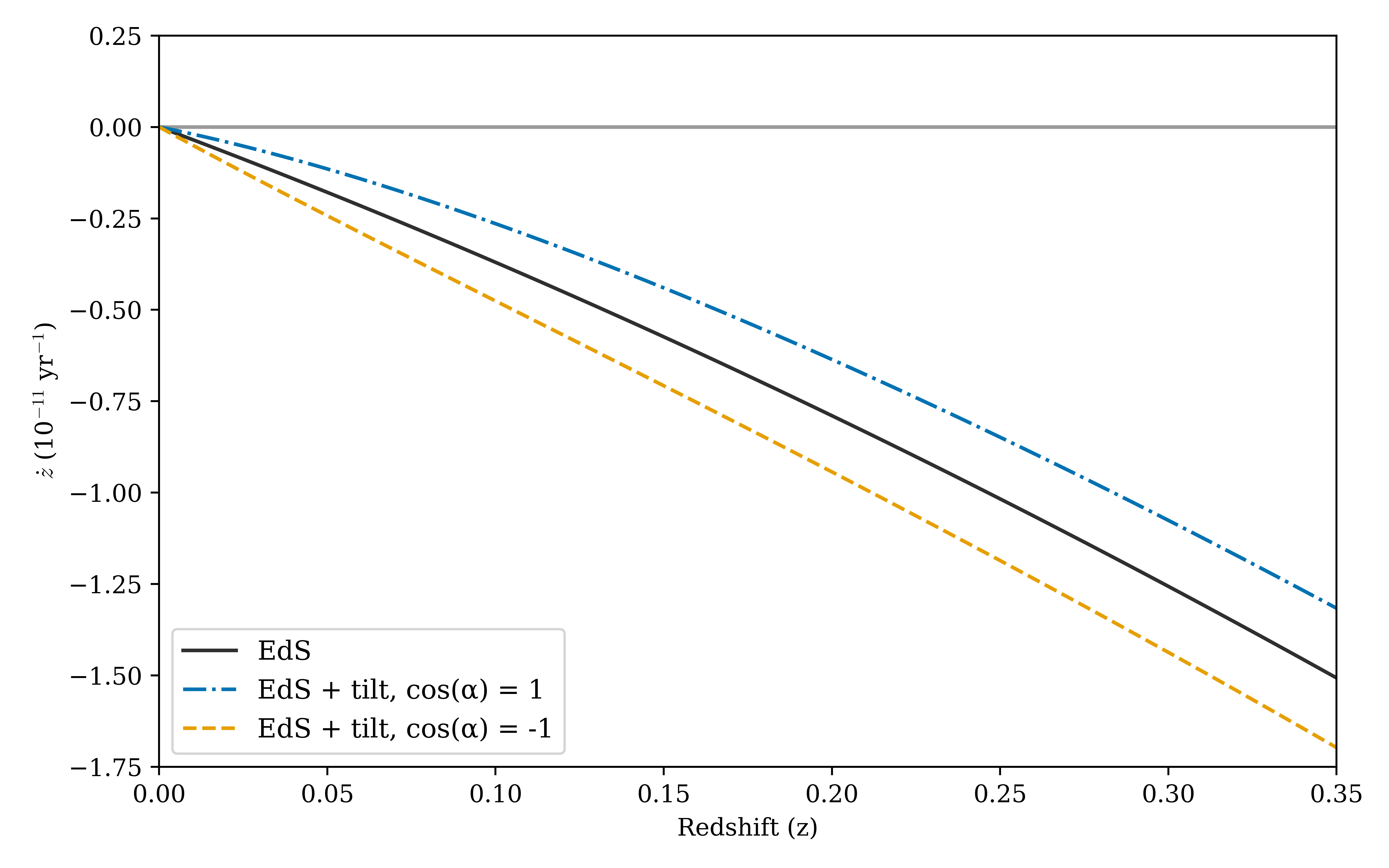}
        \caption{}
        \label{fig:EdS_low}
    \end{subfigure}
    \caption{Einstein--de Sitter redshift drift prediction with and without the CF4-normalized tilted correction. The panels compare the full-redshift behaviour with the low-redshift regime.}
    \label{EdS_results}
\end{figure}

Figure~\ref{EdS_results} shows the result of applying the CF4-normalized dipolar correction to Test A with an Einstein--de Sitter background. The solid curve corresponds to the standard EdS Sandage--Loeb signal, while the two directional curves show the maximal tilted corrections, obtained by taking \(\cos\alpha=\pm 1\). As expected, the tilted term produces an anisotropic modulation: along the direction of maximum drift the signal becomes less negative, whereas in the opposite direction it becomes more negative. However, for the observationally motivated amplitude adopted here, the correction is not large enough to reverse the sign of the EdS prediction. Thus, in this implementation, the tilted contribution distorts the EdS prediction anisotropically, but it does not mimic the positive low-redshift drift characteristic of an accelerated background. This provides a direct falsifiability criterion. If observationally motivated values of \(\delta\dot z\) cannot bring the EdS prediction close to the \(\Lambda\)CDM one, the tilted-EdS interpretation becomes testable through future redshift drift measurements.

\begin{figure}[htbp]
    \centering

    \begin{subfigure}{0.48\textwidth}
        \centering
        \includegraphics[width=\textwidth]{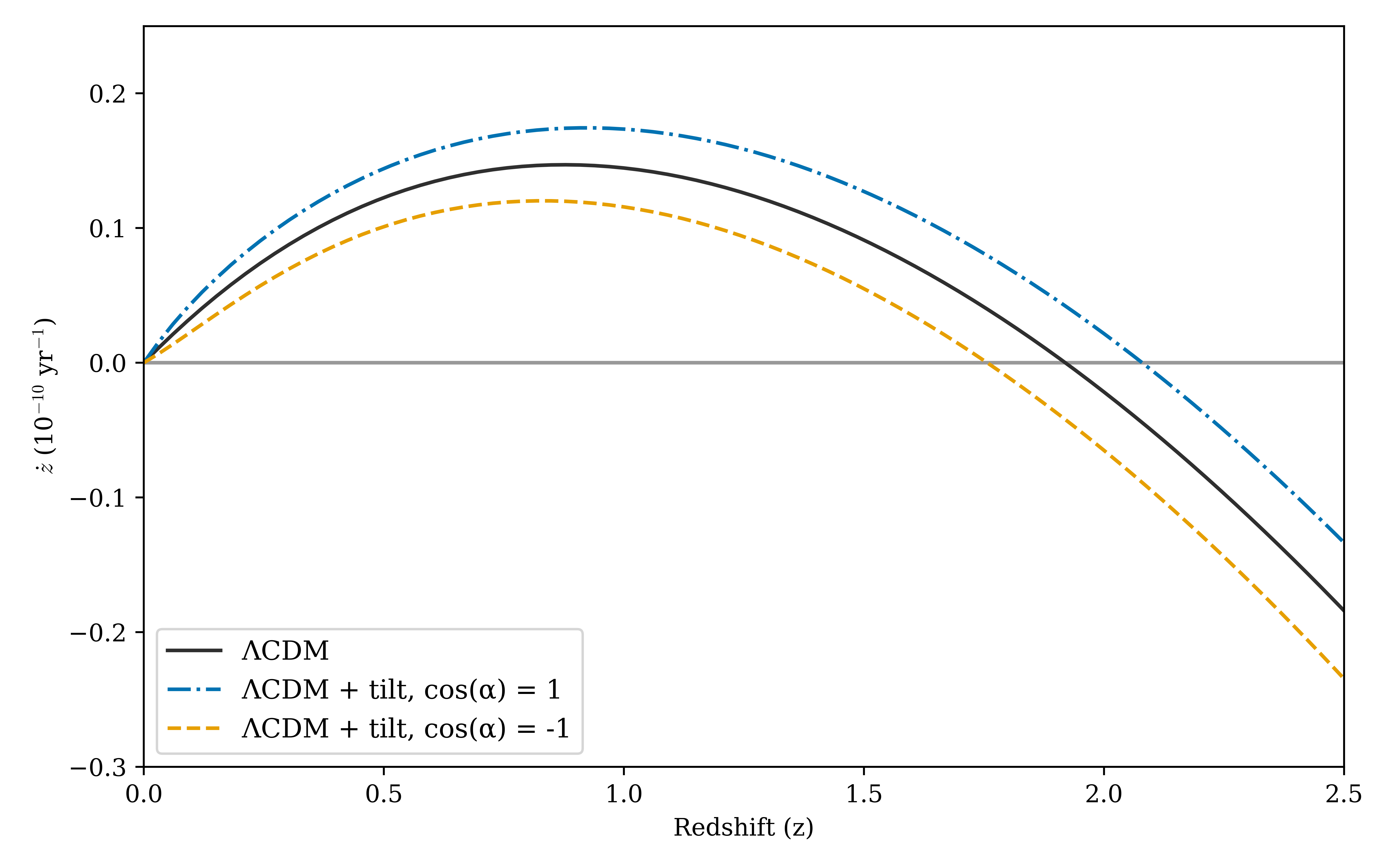}
        \caption{}
        \label{fig:LCDM_full}
    \end{subfigure}
    \hfill
    \begin{subfigure}{0.48\textwidth}
        \centering
        \includegraphics[width=\textwidth]{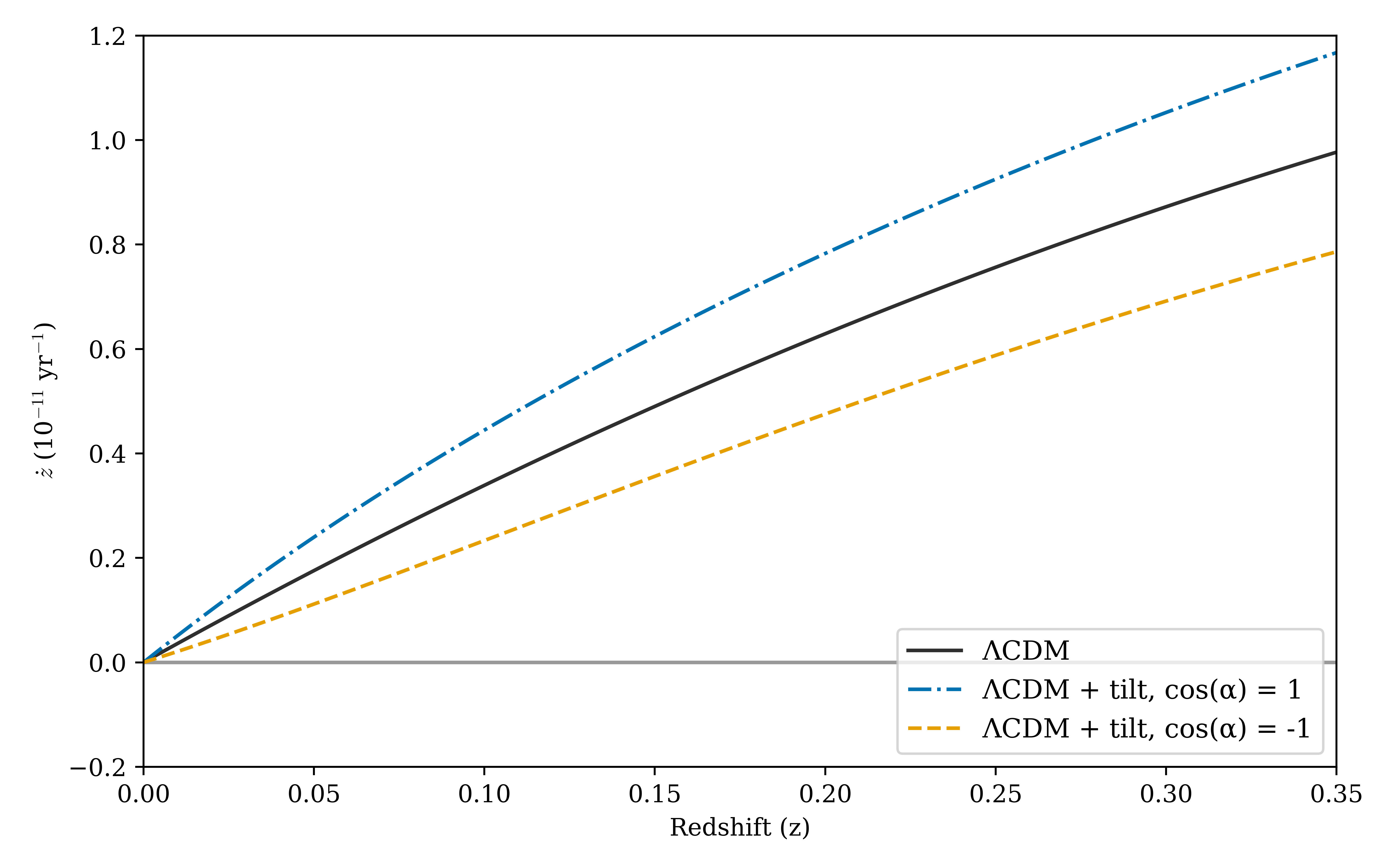}
        \caption{}
        \label{fig:LCDM_low}
    \end{subfigure}
   \caption{\(\Lambda\)CDM prediction for the redshift drift after adding the same tilted correction. The left panel shows the full-redshift range, and the right panel zooms into low redshifts.}
    \label{LCDM_results}
\end{figure}

Figure~\ref{LCDM_results} shows the corresponding prediction for Test B, when the background expansion is taken to be \(\Lambda\)CDM. In this case the positive low-redshift drift is generated by the background itself, while the tilted contribution appears as a subdominant directional modulation around the standard concordance model curve. The comparison with the EdS case therefore makes clear that the role of the tilt depends on the assumed background: in a decelerating EdS model it would have to be large enough to change the sign of the drift in order to mimic acceleration, whereas in \(\Lambda\)CDM it represents a possible anisotropic systematic superimposed on an already accelerated background. For the CF4-motivated amplitude used here, this systematic is subdominant with respect to the monopole \(\Lambda\)CDM signal, but it remains in principle falsifiable through its angular dependence and its expected alignment with the bulk flow axis.

In a strictly FLRW spacetime, the low-redshift relation \(\dot z_{\rm FLRW}=-q_0H_0z+\mathcal{O}(z^2)\) makes the sign of the redshift drift a direct probe of the background deceleration parameter. For tilted observers, however, this direct implication is lost, because the observed drift also contains the peculiar motion contribution \(\delta\dot z(\hat s,z)\). Therefore, a positive tilted redshift drift does not by itself imply global acceleration of the background universe. The criterion used here is instead model-dependent: once a background \(H(z)\) and a phenomenological model for \(\delta\dot z\) are specified, the predicted sign and angular dependence of the drift can be compared with future observations.

For the CF4-normalized tilted-EdS realization considered here, the peculiar correction is not large enough to reverse the negative EdS drift. Hence a future robust detection of a positive low-redshift drift over a broad angular domain would disfavour this specific tilted-EdS scenario as a viable replacement for dark energy, unless independently reconstructed peculiar flow gradients were found to be substantially larger than those assumed in our estimate. Conversely, the detection of a statistically significant dipolar component in \(\dot z\), especially if aligned with the observed bulk flow direction and decreasing with redshift, would provide evidence for relativistic peculiar motion effects in precision cosmology, even if such effects are not sufficient to eliminate the need for a cosmological constant.

\section{Conclusions}
\label{sec_conclusions}

In this work we derived the redshift drift measured by tilted observers within the covariant \(1+3\) formalism. Starting from the covariant definition of redshift and from the Doppler-rescaled relation between the background and tilted frames, we obtained the drift as the sum of the standard FLRW Sandage--Loeb term and a directional correction associated with the peculiar kinematics of the tilted congruence. This correction is sourced by the peculiar expansion, the projected shear, and the projected acceleration along the line of sight. In this way, the calculation keeps the background expansion \(H(z)\) explicit and treats relative motion effects as a separate light-cone correction, rather than absorbing them into effective cosmographic parameters such as \(\tilde H_{\rm eff}(z)\) or \(\tilde q(z)\).

We then implemented a minimal phenomenological model for this correction, motivated by the observed large scale bulk flow direction. Since the full peculiar velocity gradients are not yet reconstructed with sufficient precision, we normalized the dipolar correction using the CF4 bulk flow amplitude and introduced a coherence function to describe the loss of correlation with redshift beyond a transition scale. We applied this CF4-motivated dipolar model to both an Einstein--de Sitter background and the concordance \(\Lambda\)CDM model, in order to compare whether the same peculiar motion correction can mimic an accelerated drift signal in a decelerating background or instead acts only as an anisotropic systematic on top of the standard cosmology. This construction should therefore be understood as an observationally motivated estimate of the expected order of magnitude of the effect, not as a full reconstruction of the tilted redshift drift signal from peculiar velocity data.

The main result is that, for the CF4-normalized dipolar correction adopted here, the tilted contribution produces a directional modulation of the redshift drift but does not reverse the sign of the Einstein--de Sitter prediction. In the direction of maximum correction the EdS drift becomes less negative, while in the opposite direction it becomes more negative; nevertheless, \(\dot z\) remains negative throughout the low-redshift range analysed. Thus, in this concrete tilted-EdS implementation, the peculiar motion correction behaves as a subdominant anisotropic distortion of the EdS signal rather than as a mechanism capable of reproducing the positive low-redshift drift expected in \(\Lambda\)CDM.

The comparison with the \(\Lambda\)CDM case further clarifies the role of the tilted correction. If the background is already the concordance model, the positive low-redshift drift is generated by the accelerated expansion itself, and the tilted contribution appears only as a directional modulation around the standard prediction. Therefore, the same correction has two different interpretations depending on the assumed background: in EdS it would have to be large enough to mimic acceleration, whereas in \(\Lambda\)CDM it represents a possible anisotropic systematic associated with peculiar motion.

This conclusion is conditional on the phenomenological normalization used in this work. A definitive data driven prediction would require independent reconstructions of the full peculiar velocity gradients entering the covariant expression for \(\delta\dot z\), in particular \(\tilde\vartheta\), \(\tilde\sigma_{ab}\hat s^a\hat s^b\), and \(\tilde A_a\hat s^a\), together with their redshift dependence. Current reconstructions provide important evidence that the local velocity field may have nontrivial divergence and anisotropic structure, but they do not yet determine all the ingredients needed to evaluate the tilted redshift drift without additional modelling assumptions.

Consequently, once the background \(H(z)\) and the phenomenological tilted correction \(\delta\dot z(\hat s,z)\) are specified, the redshift drift provides a clean model-dependent test of tilted-EdS explanations of the apparent cosmic acceleration. In the CF4-normalized implementation considered here, a future robust detection of \(\dot z>0\) at low redshift over a wide angular domain would strongly disfavour this specific tilted-EdS realization as a replacement for dark energy, unless independently measured peculiar flow gradients were found to be substantially larger than those assumed here. Conversely, a statistically significant dipolar residual in \(\dot z\), especially one aligned with the observed bulk flow direction and decreasing with redshift, would constitute a distinctive signature of relativistic peculiar motion effects in precision cosmology, even if such effects are not sufficient to eliminate the need for a cosmological constant.

\begin{acknowledgments}

G.R.B. is supported by CONICET (Argentina) and he acknowledges support from grant PIP 112-2021-0100225-CO of CONICET (Argentina). We would like to thank Bike \& Coffee Punta Chica for the space they offer to generate and discuss ideas.

\end{acknowledgments}

\bibliography{bibliografia}

\end{document}